\documentstyle[psfig,12pt,twocolumn]{article}

\begin{document}

\title{Dialogue on the Principle of Mass-Energy Equivalence}
\renewcommand{\baselinestretch}{0.5}

\author{
{\large Ezzat G. Bakhoum}\\
\\
{\small New Jersey Institute of Technology}\\
{\small P.O. Box 305, Marlton, NJ. 08053 USA}\\
{\small Email: bakhoum@modernphysics.org}\\
\\
{\small (This work has been copyrighted with the Library of Congress)}\\
{\small Copyright \copyright 2004 by Ezzat G. Bakhoum}\\
\\
\\
\begin{minipage}{6in}
\begin{center}
{\bf Abstract}
\end{center}
{\small 
This paper is directed to the readers who are familiar with the earlier papers by the author
on the topic of mass-energy equivalence. A number of important questions about the total energy
equation $H=mv^2$ and its implications are answered qualitatively. The relationship between the
equation $H=mv^2$ and the 4-vector (Minkowski) representation of Special Relativity is discussed
in detail. Other issues, such as de Broglie's original formulation of wave mechanics, are also discussed.}
\end{minipage}
}

\date{}

\maketitle

{\large\bf Introduction:}\\
\\
This paper is a compilation of a number of important questions received by the author from several colleagues. It is organized as follows: Sec.1 contains a rather informal discussion about the implications of the total energy equation $H=mv^2$ (see ref.\cite{Bakhoum1}), presented in the form of questions and answers. The most important question asked was how a total energy equation such as $H=mv^2$ can possibly fit with the 4-vector (Minkowski) representation of special relativity, in which $H=mc^2$ is a cornerstone. This issue is addressed in detail in Sec.2. Sec.3 discusses some mathematical issues related to the concepts of the phase velocity and the group velocity, as well as de Broglie's original arguments that led to the development of wave mechanics. In Sec.4, derivations are presented for a modified Klein-Gordon equation and a de Broglie dispersion relation.

\pagebreak

{\large\bf 1. Questions and answers:}\\
\\
{\bf Q1.} According to the total energy equation $H=mv^2$, the rest energy is equal to zero. How nuclear fusion can be possible without rest energy? Two fusing particles (e.g., a deuteron and a triton) usually require little kinetic energy to start the fusion process, yet the energy released from fusion is huge!\\
\\
{\bf A.} In the fusion process, kinetic energy (usually on the order of 100Kev) is required to bring the electrostatic centers of the two fusing particles to a distance of a few fm apart, where the strong nuclear force takes over. Once the strong nuclear force takes over, the two particles move toward each other very rapidly. This relative velocity is the important parameter. The energy released in the reaction will be then equal to $\Delta m v^2$, not $\Delta m c^2$. That is why a wide distribution of energies is obtained, rather than the precise quantity $\Delta m c^2$. A detailed theory of fusion (and of fission), however, is probably still a number of years away. So, the short answer is, nuclear fusion doesn't really occur at rest.\\
\\
{\bf Q2.} At low energies, the total energy will be given by the quantity $m_0 v^2$. Hence, we have an extra mass-energy term that is equal to $\frac{1}{2} m_0v^2$ (in addition to the usual Newtonian kinetic energy). Since this energy is variable, it must be extrinsic to the particle, i.e., it can't be an intrinsic property of the particle. Where does this extra energy come from?\\
\\
{\bf A:} The answer to this question will be the Vacuum Energy, which was first proposed by Heisenberg and Euler in 1936 and verified experimentally 11 years later (see for example ref.\cite{Bordag} for the latest information on the subject). The vacuum energy has been proposed to be of the form

\[ E_{vac} = \frac{1}{2} \hbar\omega \]

What was demonstrated in ref.\cite{Bakhoum1} is that the term $\hbar\omega$ is in fact exactly equivalent to the term $m_0v^2$. This explains where the extrinsic energy $\frac{1}{2} m_0v^2$ comes from. Of course, this also suggests that the vacuum ``oscillations" (when associated with a particle) are essentially the same as de Broglie's matter waves.\\
\\
{\bf Q3.} Since $H=0$ at rest, then obviously particle decays cannot occur at rest. But some decays are actually observed to occur at rest. How can this be explained?\\
\\
{\bf A.} Care must be taken when determining whether a particle is absolutely at rest or in motion relative to the reference frame of the laboratory. For instance, it was indicated\cite{Bakhoum3} that lepton (e.g. $\mu^\pm$) and $\pi^0$ decays cannot occur at rest. On the other hand, decays such as neutron beta decay and the decay $\pi^\pm \rightarrow \mu^\pm$ can occur at rest because the decaying particle is essentially a quark that is in a state of motion within the parent meson or nucleon. Nuclear fission can also occur while the nucleus is at rest due to the phenomenon of oscillation of the fission fragments within the nucleus before the occurrence of fission. Other reactions must be interpreted in a similar manner.\\
\\
{\bf Q4.} But experiments have shown that muons ($\mu^\pm$) decay at rest.\\
\\
{\bf A.} No. In experiments such as the famous experiments of Lederman (see ref.\cite{Lederman}), muons are stopped in a block of dense material, such as carbon. It is only ``assumed" that the muon then decays at rest. There is no way to assert that the muon actually decays while at rest. In other experiments, where muons are observed to decay in photographic emulsions, there is actually conclusive evidence that the muon decays while in motion. Fig.1 below shows a typical picture of the decay chain $\pi^+ \rightarrow \mu^+ \rightarrow e^+$ in a photographic emulsion. The positive pion enters at point A in the picture and travels to point B, where it decays into a muon. The muon then travels to point C, where it decays into a positron (the positron represents the track CD). An important feature of photographic emulsions is that the thickness of the track left by the charged particle is proportional to the degree of ionization, which is in turn proportional to the velocity of the particle. In general, a slow-moving particle will produce a thicker track than a fast-moving particle. As the muon slows down in the emulsion, therefore, the track must get thicker. In the figure, if the muon had actually decelerated before decaying at point C, then the width of the track near point C must have been much larger than the width near point B. Of course, by examining the track, we can conclude that there is no noticeable difference in width along the entire segment BC. Hence, the muon decays at point C while in motion. (The distribution of the energy of the resulting positron was discussed in an earlier paper\cite{Bakhoum3}).

\par
\vspace*{.15in}

\centerline{\psfig{figure=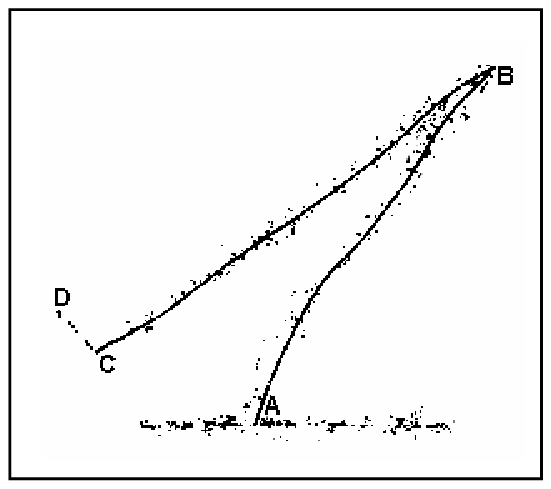}}

\vspace*{.1in}

{\small Figure 1: The decay $\pi^+ \rightarrow \mu^+ \rightarrow e^+$ in  a photographic emulsion. (Source: {\em Elementary Particles}, by I.S. Hughes, Cambridge Univ. Press. Reproduced with permission).}\\
\\
{\bf Q5.} You pointed out that the neutrino is an unnecessary hypothesis in beta decays and in lepton decays. This violates the principle of linear momentum conservation as well as angular momentum conservation!\\
\\
{\bf A.} Linear and angular momenta are not necessarily conserved in weak interactions. Specifically, since decays such as beta decays, $\mu^\pm$ decay, $\pi^\pm$ decay, $\pi^0$ decay, etc. are in most cases observed to occur inside matter, then the linear and angular momenta of the decaying particle cannot be assumed to be conserved. The momenta of nearby nuclei (or fields!) must be factored in the momentum balance. Indeed, it was pointed out\cite{Bakhoum3} that the decays $\pi^\pm \rightarrow \mu^\pm$ and  $\mu^\pm \rightarrow e^\pm$, when they occur inside matter, then they are likely one-body decays. Furthermore, when we apply the principles of energy and momentum conservation to a decay such as $\pi^0 \rightarrow \gamma\gamma$ (assuming that the energy is given by the quantity $mv^2$), then we must conclude that such a decay is a momentum non-conserving decay.\\
\\
{\bf Q6.} By saying that a neutrino does not exist in beta decay and in muon decay, are you trying to suggest that Fermi's theory of beta decay and Michel's theory of the muon decay are false?\\
\\
{\bf A.} No. Fermi's theory of beta decay and Michel's theory of muon decay are fundamentally sound theories. They successfully explained the shape of the electron's energy spectrum because they are based on the simple concept of ``momentum sharing", that is, sharing of the available momentum space between the electron and other particles. The problem is, the ``other particles" don't have to be neutrinos. As we indicated above, the other particles already exist inside matter, where the decay takes place. So, it is clear why those ``momentum sharing" theories were successful in explaining the shape of the decay spectrum. The existence of the neutrino, on the other hand, was merely a hypothesis for satisfying Einstein's equation $H=mc^2$.\\
\\
{\bf Q7.} Neutrinos from nuclear reactors have been observed experimentally. The measured neutrino flux (based on the known cross-section) also correlates very well\cite{Bemporad} with the calculated fission rate in the reactor. How can you then claim that the neutrino does not exist in beta decay?\\
\\
{\bf A.} We may very well call the highly penetrating particle that is observed to emerge from nuclear reactors a ``neutrino". However, we must point to two issues here: 1) Some preliminary calculations based on the kinematic data provided in recent reactor experiments (to be published by the author in the near future) show that this particle is actually a particle of meson-like mass and carries very little energy; 2) It is very likely that this particle emerges from the debris of particles that must be generated when the highly energetic fission products of $^{235}$U and $^{239}$Pu come to stop within the core of the reactor. This explains why the measured flux correlates with the fission rate in the reactor. The hypothesis that such a particle emerges from the beta decay of the fission products, however, is a hypothesis that cannot be verified, and, in view of the above remarks, is likely untrue.\\
\\
{\bf Q8.} The notion of ``zero rest energy" is still difficult to accept!\\
\\
{\bf A.} It may be difficult to accept, but the real question is whether this is the truth. Since a picture is usually better than a thousand words, we list below some of the known particle decay spectra (shown in the earlier papers by the author): 

\par
\vspace*{.15in}

\centerline{\psfig{figure=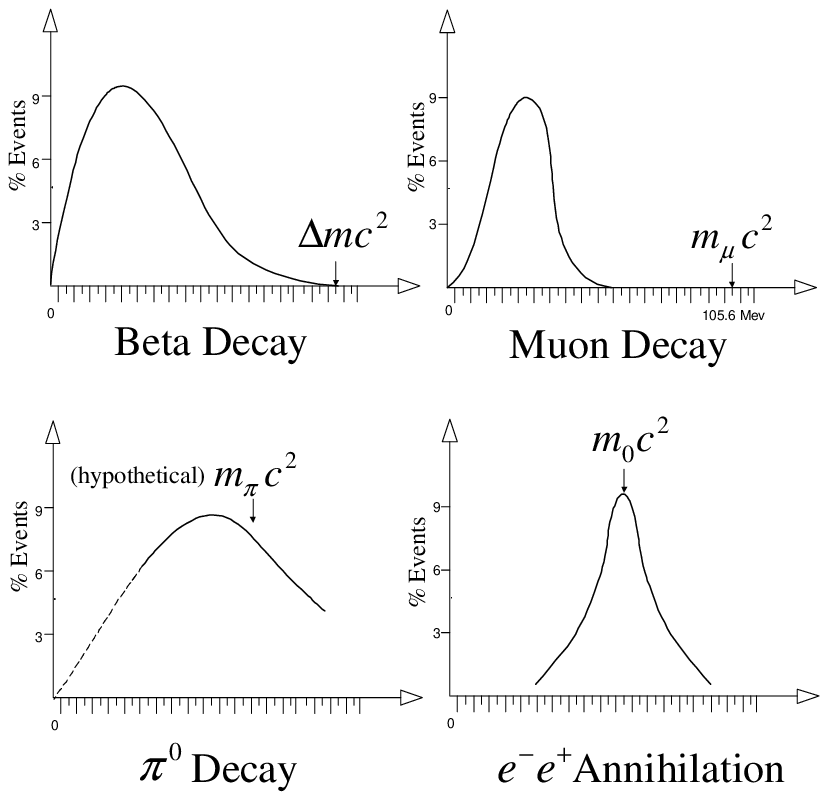}}

\vspace*{.1in}

{\small Figure 2: Various particle decay spectra. In all cases, the ``rest energy" $m_0 c^2$ either fails to emerge or appears as one data point only in a very extended spectrum.}

\pagebreak

As we can see, in many particle decay spectra, events are observed with nearly zero total energy. The appearance of the quantity $m_0 c^2$ as a peak in the annihilation spectrum of $e^-e^+$ (which also shows a Gaussian distribution of energy) was explained in ref.\cite{Bakhoum1} and suggest that the electron and the positron annihilate with a relative velocity that is approximately equal to $c$.\\
\\
{\bf Q9.} The law $H=mc^2$ may be inaccurate. But what is the assurance that $H=mv^2$ is the accurate representation of the equivalence between mass and energy?\\
\\
{\bf A.} There are a number of reasons that support the mathematical expression $H=mv^2$. They are the following: 

\begin{itemize}
\item It reconciles the two important theories of special relativity and wave mechanics (see ref.\cite{Bakhoum1}).
\item It explains all the particle decay spectra.
\item It is fully distinguishable from Bohr's theory of the atom (see ref.\cite{Bakhoum2}).
\item It explains why a reaction such as $\mu^+ \rightarrow e^+ + \gamma$ has never been observed (see ref.\cite{Bakhoum3}).
\item It predicts that the electron's wave function in the hydrogen atom must be confined within the Bohr radius, in agreement with a {\em totally independent} prediction made by Dirac (see ref.\cite{Bakhoum4}).
\end{itemize}

{\large\bf 2. On the 4-vector representation of Special Relativity:}\\
\\
In the 4-vector representation of special relativity there are often two parallel approaches that are believed to lead to the total energy equation $H=mc^2$. The first is by using the {\em invariants} of the Minkowski spacetime, specifically, the mass invariant and the length invariant; the second is by using the Lorentz transformations. Our objective in this section is to demonstrate that what applies in the case of the original formalism of Einstein is the same that applies in the case of the 4-vector formalism of Minkowski. Specifically, we will demonstrate that regardless of the approach that we choose in the Minkowski formalism, the total energy equation $H=mc^2$ was an {\em interpretation}, not a derived result.\\
\\
Another important conclusion that will emerge from this analysis is the fact that while Einstein's original formalism of 1905 was not flawed (except for the interpretative result related to the total energy), his subsequent 4-vector formalism of 1912 \cite{Einstein12} (which was clearly influenced greatly by the work of Minkowski) {\em was indeed} flawed. Specifically, the flaw was the generalization of the electromagnetic energy-momentum vector to include ordinary matter, based {\em solely} on the earlier assumption that $H=mc^2$, but based on no other physical or mathematical fact. More specifically, what we will demonstrate in the following analysis is that while the 4-velocity of a light signal is an invariant (in accordance with Einstein's postulate of 1905), the 4-velocity of a material particle {\em is not} an invariant, and hence its 4-momentum is not an invariant as the currently accepted 4-vector representation of special relativity states. Here, of course, we must point out to another serious flaw in the current understanding of many physicists: the current understanding is that the magnitude of {\em any} 4-vector in the Minkowski spacetime must be an invariant. This is incorrect for the following reason: the Minkowski formulation asserts that the ``length", or distance between any two points in spacetime, given by $(x^2+y^2+z^2 - c^2t^2)^{1/2}$ is invariant under the Lorentz transformations. This {\em does not mean}, however, that the {\em time-rate of variation} of length (a quantity that we call velocity) must also be an invariant. In fact, neither Einstein nor Minkowski suggested that a quantity such as velocity (and hence momentum) has to be invariant. Of course, this is true under the Lorentz transformations as it is true under Galilean transformations.\\
\\
An even more important result that will emerge is that if the flawed 4-velocity expression is corrected, then the total energy equation $H=mv^2$ emerges naturally! Other conclusions, such as the fact that the equation $H=mv^2$ transforms properly under the Lorentz transformations, will also be pointed to.\\
\\
{\bf 2.1. The ``invariants" of the Minkowski spacetime and the total energy formulation:}\\
\\
The idea of the {\em mass invariant} emerged from the well-known relativistic expression 

\begin{equation}
H^2 = c^2 p^2 + m_0^2 c^4
\label{21}
\end{equation}

Since this expression can be written as

\begin{equation}
- m_0^2 c^2 = p^2 - \frac{H^2}{c^2}
\label{22}
\end{equation}

Then obviously the quantity $(p^2 - H^2/c^2)$ is an invariant in any reference frame, due to the constancy of the quantity $m_0^2 c^2$. In mathematical terms, the invariant $\mbox{\boldmath $I$}_m$ is defined as 

\begin{equation}
\mbox{\boldmath $I$}_m^2 = - m_0^2 c^2 = p_\mu p^\mu,
\label{23}
\end{equation}

where $p^\mu$ is the so-called ``energy-momentum" 4-vector, defined as

\begin{equation}
p^\mu = (p_x, p_y, p_z, iH/c).
\label{24}
\end{equation}

Einstein actually {\em derived} this 4-vector for the special case of the electromagnetic field in his analysis of 1912 (the derivation was based essentially on earlier work by Poincar\'{e}) and he pointed out, in view of the above remarks about the total energy, that this form of the energy-momentum vector must be applicable to ordinary matter as well. Although we can essentially rework the same analysis by assuming that $H=mv^2$ instead of $H=mc^2$, we shall see that the problem actually has much deeper physical roots. Now, the invariant $\mbox{\boldmath $I$}_m$ obviously has a very important physical meaning. It is just the rest mass $m_0$, scaled by the universal constant $c$. Thus, the invariance of the rest mass in any reference frame is the important physical concept that we need to preserve, rather than the exact mathematical shape of the energy-momentum vector. Understanding this point is critical.\\
\\
The second invariant of the Minkowski formulation is the length invariant $\mbox{\boldmath $I$}_s$ (often called the invariant interval), defined as

\begin{eqnarray}
\mbox{\boldmath $I$}_s^2 & = & x^2 + y^2 + z^2 - c^2 t^2 \nonumber\\
      & = & x^{\prime 2} + y^{\prime 2} + z^{\prime 2} - c^2 t^{\prime 2}
\label{28}
\end{eqnarray}

This invariant, of course, is a direct consequence of the Lorentz transformations and of Einstein's postulate of the constancy of the speed of light $c$ in all frames of reference. $\mbox{\boldmath $I$}_s$ can then be written as a 4-vector

\begin{eqnarray}
\mbox{\boldmath $I$}_s & = & (x, y, z, ict), \quad \mbox{in $S$, or}\nonumber\\
\mbox{\boldmath $I$}_s & = & (x^\prime, y^\prime, z^\prime, ict^\prime) \quad \mbox{in $S^\prime$}
\label{29}
\end{eqnarray}

Here, we are making the usual assumption that a certain reference frame $S$ is stationary, while a particle in its own rest frame $S^\prime$ is moving with a constant velocity $v$ with respect to $S$. Of course, $\mbox{\boldmath $I$}_s$ in either case is a representation of the Minkowski coordinates. By differentiating $\mbox{\boldmath $I$}_s$ with respect to the proper time $\tau$ of the particle, we obtain the so-called ``4-velocity" $\vec{u}$ of the particle (where, as usual: $d\tau \equiv dt^\prime = dt/\gamma$, $\gamma = (1-v^2/c^2)^{-1/2}$ and $dx/d\tau = (dx/dt).(dt/d\tau) = \gamma v_x$); the 4-velocity is then given by

\begin{eqnarray}
\vec{u} & = & (\gamma v_x, \gamma v_y, \gamma v_z, i \gamma c), \quad \mbox{in $S$, or}\nonumber\\
\vec{u^\prime} & = & (0, 0, 0, ic) \quad \mbox{in $S^\prime$}
\label{210}
\end{eqnarray}

It is now easy to see (by virtue of the fact that $\gamma^2 = (1-v^2/c^2)^{-1}$ and $v_x^2 + v_y^2 + v_z^2 = v^2$) that 

\begin{equation}
(u)^2 = (u^\prime)^2 = - c^2.
\label{211}
\end{equation}

A common ``derivation" of the mass-energy equivalence relationship (not originally given by Einstein) is to then multiply the mass invariant $m_0$ by this new ``velocity invariant", $c^2$, and hence obtain the famous equation $H_0 = m_0 c^2$. Why this particular product should be the correct relationship between mass and energy? (as the reader will now realize, Eq.(\ref{211}) is {\em merely} a statement of the principle of the constancy of the velocity of light in all reference frames. The notion of ``energy" is not present at all in the above relationships). The issue goes back to Einstein's old interpretation. Such a product is just what Einstein subconsciously predicted as the equivalence between mass and energy.\\
\\
We can then see that the above derivation of the famous formula is essentially nothing but Einstein's old interpretation. But this is not really the important issue here. To see what the important issue is, let us look again at the meaningless result in Eq.(\ref{210}). This result will be indeed meaningless if we interpret it as the ``velocity of a particle", because, in such a case, it must be taken to mean that ``the velocity $u^\prime$ of the particle in its own rest frame is equal to $c$". Of course, we are talking about a 4-dimensional space here rather than the usual 3-dimensional space; but what does a result like that mean {\em physically}? As we must now fully realize, Eqs.(\ref{210}) and (\ref{211}) are essentially statements of the principle of the constancy of the velocity of light $c$. In words, they can be written as follows: ``The velocity of light $c$ is the same in the particle's frame as it is in any other frame". Clearly then, we {\em cannot} make the conclusion that $c$ is one of the velocity components of the particle. Let us look at the problem from a much wider perspective: this kind of problem actually has roots which are deeply embedded in the subconscious of physicists when they use the powerful tool of mathematics. When we deal with a particle or a system of  particles, we must understand that we are dealing essentially with two different physical problems: the first problem is the problem of kinematics and the second problem is the coordination of such kinematics by means of light signals. Although the two problems must obviously be related in any physical theory, there is actually a huge difference between the two. Concepts such as velocity, momentum and energy are usually related to the problem of kinematics, while concepts such as causality are usually related to the problem of coordination. The lack of understanding of this difference can lead to serious flaws in any theory (there is no question that Einstein's analysis in his manuscript of 1912 \cite{Einstein12} was {\em greatly} influenced by the purely mathematical ideas of Minkowski). Let us examine more carefully what the Minkowski coordinates in Eq.(\ref{29}) mean. Those coordinates, of course, resulted from the invariant representation of Eq.(\ref{28}). This invariant, $\mbox{\boldmath $I$}_s^2$, is essentially more or less a statement of the principle of the constancy of the velocity of light and the fact that the coordinates $(x, y, z)$ and $(x^\prime, y^\prime, z^\prime)$ satisfy the Lorentz transformations. Hence, the Minkowski coordinates are coordinates that were derived purely from the problem of the coordination of events by means of light signals. This is why the velocity of light $c$ appears in the 4th component of the vector $(x^\prime, y^\prime, z^\prime, ict^\prime)$. While the 4th coordinate $ct^\prime$ is essential for the Minkowski formalism (because it is based on the Lorentz transformations) we must be very careful when we decide to take the time derivative of that 4th coordinate in order to calculate a velocity component for a particle, because we will then be mixing the problem of the coordination of events with the problem of the particle's kinematics, which is an approach that surely leads to meaningless results. As we saw, the differentiation led to a velocity component equal to $c$, which is just the velocity of light that was needed in order to solve the coordination problem. As a hypothetical example, if we had used $10^6 ct^\prime$ instead of $ct^\prime$ as the 4th coordinate, we would have obtained a velocity component equal to $10^6 c$. Does this then mean that the final calculation of the energy of the particle will be correct? How about a purely Galilean transformation in which $c=\infty$, what does this tell us then about the ``4th component" of the velocity $u^\prime$ of the particle? We can therefore clearly see the danger of taking the concepts which are used solely to solve the problem of the coordination of events and using them to predict numerical values for the dynamical properties of a particle.\\
\\
How can we solve the above problem, then, and still maintain the 4-vector formalism of Minkowski? The solution is to separate the kinematics problem from the coordination problem. Specifically, we must assume that the 4-velocity $u^\prime$ of the particle in its own rest frame is the same as its 3-velocity: 0. Hence, we must assume that $u^\prime = (0,0,0,0)$. This means that when we take the derivative of the 4th Minkowski coordinate $ict^\prime$ (in Eq.{\ref{29}) with respect to the proper time $\tau$, the result must be 0. Hence, we must assume that (note that $\tau \equiv t^\prime$)

\begin{equation}
\frac{d}{d\tau} (ct^\prime) = 0 = c + t^\prime \frac{dc}{d\tau}
\label{212}
\end{equation}

This means that

\begin{equation}
\frac{dc}{d\tau} = - \frac{c}{t^\prime} 
\label{213}
\end{equation}

At a first glance, it may seem that we are taking a potentially wrong approach, since we are obviously violating the principle of the constancy of the velocity of light $c$! Not so! When the particle under consideration is a material particle, then the $c$ in the Minkowski coordinate $ict^\prime$ is not usually interpreted as the ``velocity of light", rather, in relativistic language, it is referred to as the ``light-like" velocity component of the particle. In fact, it is purely a mathematical abstraction. This must be especially clear in view of the above discussion. There is absolutely no reason, on either physical or mathematical grounds, that this ``light-like" component must be treated as a constant. This was the flaw in Einstein's analysis of 1912. As we shall see, when this restriction is removed, remarkable results will be indeed obtained. Now, what Eq.(\ref{213}) means is that the ``light-like" component of the velocity will be decelerating in the particle's rest frame, so the net result is that the ``kinematic", or physical velocity of the particle is zero.\\
\\
In order to obtain the 4-velocity of the particle in the observer's frame $S$, we differentiate the Minkowski coordinates in Eq.(\ref{29}) again with respect to the proper time $\tau$. The first three components will remain unchanged (that is, $\gamma v_x, \gamma v_y, \gamma v_z$). The 4th component will now give (note that $t^\prime = \gamma t$ if we let $x=0$ in the Lorentz transformation)

\begin{eqnarray}
\frac{d}{d\tau} (ct) & = & c \frac{dt}{d\tau} + t \frac{dc}{d\tau}\nonumber\\
                     & = & \gamma c + t \left(\frac{-c}{\gamma t} \right)\nonumber\\
                     & = & \gamma c - \frac{c}{\gamma}\nonumber\\
                     & = & \gamma c (1 - \frac{1}{\gamma^2})\nonumber\\
                     & = & \gamma c (1- (1- v^2/c^2))\nonumber\\
                     & = & \frac{\gamma v^2}{c}
\label{214}
\end{eqnarray}

The 4-velocity $\vec{u}$ of the particle will then be

\begin{equation}
\vec{u} = (\gamma v_x, \gamma v_y, \gamma v_z, i \gamma \frac{v^2}{c})
\label{215}
\end{equation}

and hence $u^2$ is

\begin{eqnarray}
u^2 & = & \gamma^2 v^2 - \gamma^2 \frac{v^4}{c^2}\nonumber\\
    & = & \gamma^2 v^2 \left(1 - \frac{v^2}{c^2} \right)\nonumber\\
    & = & v^2
\label{216}
\end{eqnarray}

Now we note that in the special case of $v=c$, $\vec{u}$ in Eq.(\ref{215}) becomes $(\gamma v_x, \gamma v_y, \gamma v_z, i \gamma c)$ and hence $u^2 = - c^2$ again. Accordingly, $u^2$ is an invariant {\em only} in the special case of a light signal.

\pagebreak

What do the above results then mean as far as the concept of the total energy of a material particle is concerned? The answer is very clear: if we follow the traditional approach, then $H_0 = m_0 u^{\prime 2} = 0$ in the particle's rest frame and $H = m u^2 = m v^2$ in the observer's frame. Let us now obtain an expression for the 4-momentum. The 4-momentum of the particle can be obtained by simply multiplying the 4-velocity in Eq.(\ref{215}) by $m_0$. Hence

\begin{equation}
p^\mu = (\gamma m_0 v_x, \gamma m_0 v_y, \gamma m_0 v_z, i \gamma m_0 \frac{v^2}{c})
\label{218}
\end{equation}

Compare this expression for the 4-momentum with the original expression in Eq.(\ref{24}). The only difference is the appearance of the total energy $mv^2$ in the last term instead of the total energy $mc^2$. It is now straightforward to verify that $p_\mu p^\mu$, or the square of the 4-momentum vector, is equal to $m_0^2 v^2$. This again is to be compared with the original expression $p_\mu p^\mu = - m_0^2 c^2$. Here, we must of course emphasize the fact that, contrary to the popular belief, quantities such as velocity and momentum {\em are not} universally invariant under the formulation of Minkowski (they are of course invariant in any {\em single} frame of reference, as the laws of conservation of energy and momentum require). An additional important fact about the two expressions $p_\mu p^\mu = - m_0^2 c^2$ and  $p_\mu p^\mu = m_0^2 v^2$ is their physical meaning. If we ask whether these expressions represent momentum or represent energy, the answer is neither. This is because the vector $p^\mu$ itself is a ``hybrid" vector of energy and momentum. Traditionally, the old expression $p_\mu p^\mu = - m_0^2 c^2$ has been applied in problems such as particle collisions and particle decays and was assumed to represent the ``momentum squared". In practice, it merely leads to a mass conservation law. Because of the mathematical nature of the 4-vector $p^\mu$, such an expression {\em cannot} be understood as a ``momentum squared". When we write momentum conservation and energy conservation equations in cases like particle collisions or particle decays, we must deal with the components of the 4-vector $p^\mu$ in Eq.(\ref{218}) {\em separately}. The first three components are a true representation of momentum, while the 4th component is a true representation of energy. Hence, a quantity such as $p_\mu p^\mu$ cannot be used by itself in a conservation law, since there is no ``hybrid" conservation law that comprises both energy and momentum.\\
\\
To summarize the above discussion: according to the modified formulation, the rest mass $m_0$ and the spacetime interval $\mbox{\boldmath $I$}_s^2$ are invariants as in the usual theory, but the 4-velocity and the 4-momentum {\em are not}. They are invariants {\em only} in the special case of a light signal.\\ 
\\
{\bf 2.2. The connection between the total energy and the Lorentz transformations:}\\
\\
In view of the conclusions just reached, we shall now proceed to investigate the relationship between the total energy and the Lorentz transformations. The relationship between the total energy and the ``plain" Lorentz transformations (that is, excluding the Minkowski formulation) is discussed in the Appendix to this paper.\\
\\
If we assume the simple case of motion along one axis only, the Lorentz transformations are

\begin{eqnarray}
x^\prime & = & \gamma (x - vt) \nonumber\\
y^\prime & = & y \nonumber\\
z^\prime & = & z \nonumber\\
t^\prime & = & \gamma (t - \frac{vx}{c^2}),
\label{219}
\end{eqnarray}

or, in matrix form (and using the Minkowski coordinates),

\begin{equation}
\left( \begin{array}{c}
x^\prime\\
y^\prime\\
z^\prime\\
ict^\prime
\end{array} \right)
=
\left( \begin{array}{cccc}
\gamma & 0 & 0 & i \gamma v/c\\
0 & 1 & 0 & 0\\
0 & 0 & 1 & 0\\
-i \gamma v/c & 0 & 0 & \gamma
\end{array} \right)
\left( \begin{array}{c}
x\\
y\\
z\\
ict
\end{array} \right)
\label{220}
\end{equation}

When we differentiate the coordinates with respect to the proper time $\tau$, we note that we must take the derivative of any coefficient to the coordinate $ict$, because the variable ``$c$" (when it is attached to that coordinate) must be treated as the ``light-like" component discussed above. In the above transformation matrix, the coefficients of $ict$ are the quantities $i \gamma v/c$ in the first row, and $\gamma$ in the last row. Since $\gamma$ is a constant, then $d\gamma /d\tau =0$. The derivative of $i \gamma v/c$ is $i \gamma v . d/d\tau (c^{-1})$, where $c$ is the light-like component. It is straightforward to verify that this quantity is equal to $i \gamma v / c t^\prime = i \gamma v / c \gamma t = i v / ct$. Hence, when we differentiate with respect to $\tau$, the above matrix representation will be written as follows (note that the derivative of $ict^\prime$ is equal to 0 and that of $ict$ is equal to $i\gamma v^2/c$):

\begin{eqnarray}
\lefteqn{\left( \begin{array}{r}
v_x^\prime =0\\
0\\
0\\
0
\end{array} \right) =} & \nonumber\\
 & \left( \begin{array}{cccc}
\gamma & 0 & 0 & i \gamma v/c\\
0 & 1 & 0 & 0\\
0 & 0 & 1 & 0\\
-i \gamma v/c & 0 & 0 & \gamma
\end{array} \right)
\left( \begin{array}{c}
\gamma v_x\\
0\\
0\\
i \gamma v^2/c
\end{array} \right) & \nonumber\\
 & + \left( \begin{array}{cccc}
0 & 0 & 0 & iv/ct\\
0 & 0 & 0 & 0\\
0 & 0 & 0 & 0\\
0 & 0 & 0 & 0
\end{array} \right)
\left( \begin{array}{c}
x\\
y\\
z\\
ict
\end{array} \right),
\label{221}
\end{eqnarray}

which can also be written as

\begin{eqnarray}
\lefteqn{\left( \begin{array}{r}
v_x^\prime =0\\
0\\
0\\
0
\end{array} \right) =} & \nonumber\\
 & \left( \begin{array}{cccc}
\gamma & 0 & 0 & i \gamma v/c\\
0 & 1 & 0 & 0\\
0 & 0 & 1 & 0\\
-i \gamma v/c & 0 & 0 & \gamma
\end{array} \right)
\left( \begin{array}{c}
\gamma v_x\\
0\\
0\\
i \gamma v^2/c
\end{array} \right) & \nonumber\\
 & - \left( \begin{array}{c}
v\\
0\\
0\\
0
\end{array} \right)
\label{222}
\end{eqnarray}

The first of the above equations will be $v_x^\prime = 0 = \gamma(\gamma v_x) - (\gamma v/c)(\gamma v^2/c) - v$. It is again straightforward to prove that the sum $(\gamma v/c)(\gamma v^2/c) + v$ of the last two terms reduces to $(\gamma c/v)(\gamma v^2/c)$. Hence the above set of equations can be written as

\begin{eqnarray}
\lefteqn{\left( \begin{array}{r}
v_x^\prime =0\\
0\\
0\\
0
\end{array} \right) =} & \nonumber\\
 & \left( \begin{array}{cccc}
\gamma & 0 & 0 & i \gamma c/v\\
0 & 1 & 0 & 0\\
0 & 0 & 1 & 0\\
-i \gamma v/c & 0 & 0 & \gamma
\end{array} \right)
\left( \begin{array}{c}
\gamma v_x\\
0\\
0\\
i \gamma v^2/c
\end{array} \right)   
\label{223}
\end{eqnarray}

If we finally multiply each of the equations by $m_0$, we obtain the vector $p^\mu$ of Eq.(\ref{218}) on the r.h.s., that is,

\begin{equation}
\left( \begin{array}{r}
0\\
0\\
0\\
0
\end{array} \right) =
\left( \begin{array}{cccc}
\gamma & 0 & 0 & i \frac{\gamma c}{v}\\
0 & 1 & 0 & 0\\
0 & 0 & 1 & 0\\
-i \frac{\gamma v}{c} & 0 & 0 & \gamma
\end{array} \right)
\left( \begin{array}{c}
\gamma m_0 v_x\\
0\\
0\\
i \gamma m_0 v^2/c
\end{array} \right)
\label{224}
\end{equation}

The question now, of course, is whether each of the components of the $p^\mu$ vector (including the total energy $\gamma m_0 v^2$ in the last component) identically transforms to 0 in the particle's frame. Since $v_x = v$, then a simple check shows that this is indeed the case. The conclusion is hence that {\em both} the energy $mv^2$ and the momentum $mv$ transform properly under the Lorentz transformations to a null quantity in the particle's frame. In the Appendix, a connection is made between the total energy $mv^2$ and the ``plain" Lorentz transformations.\\
\\
\\
{\large\bf 3. Issues related to the original formulation of wave mechanics:}\\
\\
{\bf 3.1. The concepts of phase velocity and group velocity:}\\
\\
One important question that was asked repeatedly after the publication of the first paper on this subject\cite{Bakhoum1} is why the phase velocity of a matter wave cannot be equal to $c^2/v$ as the original theory predicts. We will attempt in this section to give a better explanation for that problem.\\
\\
First of all, the concept itself was mathematically flawed. Superposition of waves does not result in {\em one distinct} phase velocity. We must have a number of phase velocities. Now, when we superimpose waves, it is well known mathematically that each phase velocity $v_p = \omega_i / k_i$ and that the group velocity $v_g = d\omega /dk$ (where $\omega$ is the angular frequency and $k=2\pi / \lambda$ is the propagation constant). As was pointed out in ref.\cite{Bakhoum1}, the two fundamental relationships of wave mechanics: $\lambda = h/p$ and $H=h\nu$ together make a {\em statement} about the total energy of a particle; that is, $H=(p\lambda)\nu=pu$, where $u$ is some velocity. The question here is what is $u$ exactly? Is it a phase velocity or a group velocity? Apart from the fact that $H=pu$ is a {\em total} energy equation, we must also note, since $H=\hbar\omega$ and $p=\hbar k$, that the equation leads to the relationship $\omega = k u$. Hence we must conclude that 

\begin{equation}
\frac{d\omega}{dk} = \frac{\omega}{k} = u
\label{41}
\end{equation}

This means that the group and the phase velocities are {\em the same}. Note that in arriving to the conclusion in Eq.(\ref{41}) we have of course assumed that $u$ is a constant, i.e., not a function of $k$. Indeed, the traditional approach to the problem is to assume that $u$ has a value of $c^2/v$ (a constant), where $v$ is the particle's velocity, in order to satisfy the relativistic equation $H=mc^2$ (as was pointed out in ref.\cite{Bakhoum1}). $u$ is then called the ``phase velocity". But, in order to satisfy the basic principles of physics, another assumption that also holds is that $d\omega / dk$, or the group velocity, must be equal to $v$. Hence it is clear, in view of Eq.(\ref{41}), that the fundamental inconsistency between wave mechanics and special relativity has never been resolved. Of course, a quantity that is taken by default to be equal to $c^2/v$ {\em cannot} also be equal to $v$. Simply stated, the velocity of propagation of energy cannot have two different values.\\
\\
One important derivation that is given by many authors and which ``seems" to bypass the problem (see for example ref.\cite{French}) starts with the longer relativistic expression $H^2 = p^2c^2 + m_0^2 c^4$, calculates an expression for the group velocity $d\omega / dk$ as a function of $\omega / k$, and since $d\omega / dk$ must be equal to $v$, the result that is reached then is that  $\omega/k = v_p = c^2/v$. This derivation, however, is no different from the above derivation in that, by ``forcing" the group and the phase velocities to have different values, it {\em completely} ignores the simple and unequivocal mathematical relationship in Eq.(\ref{41}). As was demonstrated in ref.\cite{Bakhoum1}, if we abide by Eq.(\ref{41}), that is, assume that $u=v_g=v_p=v$, then we reach the conclusion that the total energy $H=pu=mv^2$. In Sec.4 of this paper, it is shown that if we use the longer relativistic expression $H^2 = p^2c^2 - m_0^2 v^2c^2$ (where $H=mv^2$ in this modified equation), we reach again the result that $v_g=v_p=v$.\\
\\
{\bf 3.2. De Broglie's derivation of the relationship $\lambda = h/p$}:\\
\\
De Broglie's original derivation of the important relationship $\lambda = h/p$ can be found in a number of standard references (see ref.\cite{French} for example). Amazingly, as we shall conclude, while the formula was correct, the approach that was used to derive it {\em was not}.\\
\\
De Broglie started by assuming a wave function that describes a stationary particle, of the form $\psi^\prime = \exp (i \omega^\prime t^\prime)$. By using the Lorentz transformation of time $t^\prime = \gamma (t - vx/c^2)$, then $\psi^\prime = \exp (i \gamma \omega^\prime [t - vx/c^2])$. Since this equation (in principle) is a traveling wave equation, de Broglie then concluded that the quantity $c^2/v$ must represent the velocity of the wave in the observer's frame. The rest of the derivation that leads to the formula $\lambda = h/p$ is then straightforward and consists of letting $H=h\nu=mc^2$ and substituting the product $\lambda\nu$ for the quantity $c^2/v$. As it is well known historically\cite{deBroglie}, de Broglie later offered the hypothesis that $c^2/v$ is only a ``phase" velocity and that the real, or ``group" velocity is actually $v$ so that the particle and its associated wave would not part company. However, as we indicated, the problem with such a hypothesis is that it directly contradicts the simple conclusion in Eq.(\ref{41}).\\
\\
Let us try to understand the problem with the above approach that led to the indicated contradictions. The Lorentz transformation of time $t^\prime = \gamma (t - vx/c^2)$, which includes the coordinate $x$, strictly assumes that ``$x$" is only one geometrical point. From the viewpoint of a stationary observer, a traveling wave, in the observer's frame, cannot be described by one ``$x$" coordinate. The correct approach for including a traveling wave within the relativistic transformations is to assume first that the ``$x$" coordinate is equal to zero (and hence the time transformation will be $t^\prime = \gamma t$) and then write a {\em true} traveling wave equation in the observer's frame, that is

\begin{equation}
\psi = \exp i(kx-\omega t)
\label{42}
\end{equation}

This was indeed the approach that was taken by Shr\"{o}dinger and certainly this explains why Shr\"{o}dinger's equation has been unquestionably successful (there is no doubt that de Broglie, as a Ph.D. student, was under pressure to formulate his theory in a manner that does not lead to contradictions with $H=mc^2$). Now, by noting that $k=2\pi / \lambda$ and $\omega = 2\pi\nu$, $\psi$ can be written as

\begin{eqnarray}
\psi & = & \exp i \left(\frac{2\pi}{\lambda} x-\omega t \right)\nonumber\\
     & = & \exp i \omega \left( \frac{2\pi}{\lambda} \frac{x}{2\pi\nu} - t \right)\nonumber\\
     & = & \exp i \omega \left(\frac{x}{\lambda\nu} - t \right)
\label{43}
\end{eqnarray}

Assume first that the particle is moving with a velocity $v\ll c$ so that the relativistic effects can be ignored. In this case, ordinary (non relativistic) wave mechanics state that $\lambda\nu = v$, or the wave's velocity. Now, if the relativistic effect is to be included, then the wavelength $\lambda$ becomes $\lambda / \gamma$ (length contraction) and the frequency $\nu$ becomes $\gamma\nu$ (frequency shift). The result therefore is that $\lambda\nu$ is still equal to $v$. We can see, then, that the flaw in the original approach that led to the result $\lambda\nu = c^2/v$ was the incorrect use of the Lorentz transformation.\\
\\
If we now follow the rest of de Broglie's derivation, but use $H=mv^2$ instead of $mc^2$, we have $H=mv^2=h\nu$, hence

\begin{eqnarray}
p = mv & = & \frac{h\nu}{v}\nonumber\\
       & = & \frac{h\nu}{\lambda\nu}\nonumber\\
       & = & \frac{h}{\lambda}
\end{eqnarray}

Which is of course de Broglie's well known formula.  De Broglie was aware that this relationship can be derived in a number of different ways, and for that reason he raised it to the level of a {\em postulate}. Concerning the approach that was used in deriving it, however, this is certainly one of the rare cases in science in which an incorrect derivation procedure still led to the correct result.\\
\\
\\
{\large\bf 4. A modified Klein-Gordon equation and de Broglie dispersion relation:}\\
\\
In this section we present derivations for a modified Klein-Gordon equation and a modified de Broglie dispersion relation. The conclusions are: 1) in the case of a massless particle, the dispersion relation is the same as the original one; and 2) in the case of a massive particle, we still conclude that the phase and the group velocities are the same, that, is $v_g = v_p = v$.\\
\\
{\bf 4.1. The Klein-Gordon equation:}\\
\\
The derivation of the Klein-Gordon equation starts with the usual relativistic expression (see ref.\cite{Griffiths})

\begin{equation}
H^2 = p^2 c^2 + m_0^2 c^4
\label{31}
\end{equation}

If we now replace $H$ by $mc^2$ and $p$ by $mv$ we have

\begin{equation}
m^2 c^4 = m^2 v^2 c^2 + m_0^2 c^4
\label{32}
\end{equation}

If we multiply this expression by $v^2/c^2$, we get

\begin{equation}
m^2 v^2 c^2 = m^2 v^4 + m_0^2 v^2 c^2
\label{33}
\end{equation}

If we now let $H=mv^2$ we finally have

\begin{equation}
H^2 = p^2 c^2 - m_0^2 v^2 c^2
\label{34}
\end{equation}

This is a modified energy-momentum relationship and was in fact derived previously in ref.\cite{Bakhoum1}. Notice that the quantity $m_0^2 v^2 = p^2 - H^2/c^2$. It is therefore a correct representation of the vector $p^\mu$ in Eq.(\ref{218}).\\
\\
To obtain the modified Klein-Gordon equation, we start with the well known relationship

\begin{equation}
\nabla^2 \psi =  - \mbox{\boldmath $k$}^2 \psi = - \frac{\mbox{\boldmath $p$}^2}{\hbar^2} \psi
\label{35}
\end{equation}

By substituting from Eq.(\ref{34}) into Eq.(\ref{35}) we have

\begin{equation}
- \hbar^2 \nabla^2 \psi = \left(\frac{H^2}{c^2} + m_0^2 v^2 \right) \psi
\label{36}
\end{equation}

From Shr\"{o}dinger's equation we have

\begin{equation}
\frac{\partial^2 \psi}{\partial t^2} = - \frac{H^2}{\hbar^2} \psi
\label{37}
\end{equation}

By substituting from Eq.(\ref{36}) into Eq.(\ref{37}) we finally get

\begin{equation}
\frac{1}{c^2} \frac{\partial^2 \psi}{\partial t^2} - \nabla^2 \psi = \left(\frac{m_0 v}{\hbar}\right)^2 \psi 
\label{38}
\end{equation}

This is the modified Klein-Gordon equation.\\
\\
{\bf 4.2. De Broglie's dispersion relation:}\\
\\
In view of Eqs.(\ref{35}) and (\ref{37}), the modified Klein-Gordon equation can be written as

\begin{eqnarray}
- \frac{1}{c^2} \left(\frac{\omega^2 \hbar^2}{\hbar^2}\right) \psi & = & - \mbox{\boldmath $k$}^2 \psi + \left(\frac{m_0 v}{\hbar}\right)^2 \psi, \quad \mbox{or,}\nonumber\\
\omega^2 \hbar^2 \psi & = & c^2 \hbar^2 \mbox{\boldmath $k$}^2 \psi - m_0^2 c^2 v^2 \psi
\label{39}
\end{eqnarray}

and hence the modified de Broglie wave dispersion relation is

\begin{equation}
\hbar^2\omega^2 = c^2 \hbar^2 \mbox{\boldmath $k$}^2 - m_0^2 c^2 v^2 
\label{310}
\end{equation}

For $m_0=0$, we can see that the relation becomes $\hbar^2\omega^2 = c^2 \hbar^2 \mbox{\boldmath $k$}^2$, which is of course the same as in the usual theory.\\
\\
To obtain the group velocity, $\mbox{\boldmath $v$}_g = d\omega / d \mbox{\boldmath $k$}$, we differentiate the dispersion relation with respect to \mbox{\boldmath $k$}, getting (note: only the magnitudes of the vectors \mbox{\boldmath $p$} and \mbox{\boldmath $k$} will be represented)

\begin{equation}
\hbar^2\omega \frac{d \omega}{d k}= c^2 \hbar^2 k - m_0^2 c^2 v \frac{d v}{d k}
\label{311}
\end{equation}

Since $p=mv=\hbar k$ and hence $m (dv/ dk) = \hbar$, the above equation becomes

\begin{equation}
\hbar^2\omega \frac{d \omega}{d k}= c^2 \hbar^2 k - m_0^2 c^2 \frac{\hbar^2}{m^2} k
\label{312}
\end{equation}

or,

\begin{eqnarray}
\omega \frac{d \omega}{d k} & = & c^2 k - \frac{m_0^2}{m^2} c^2 k\nonumber\\
  & = & c^2 k (1-(1-\frac{v^2}{c^2}))\nonumber\\
  & = & k v^2
\label{313}
\end{eqnarray}

Hence

\begin{equation}
\frac{d \omega}{d k} = \left(\frac{k}{\omega} \right) v^2 
\label{314}
\end{equation}

But since $d\omega / dk = v_g = v$, we then conclude that $\omega / k = v_p = v$. The group and the phase velocities are therefore the same.\\
\\
\\
{\large\bf Summary:}\\
\\
From this paper and the first paper by the author on this subject\cite{Bakhoum1} it must be clear that the fundamental inconsistency between the theories of special relativity and wave mechanics had not been resolved, despite de Broglie's efforts to reconcile the two theories. This fact, together with the unequivocal failure of the equation $H=mc^2$ to explain many experimental (and theoretical) results, point to one inevitable conclusion: Einstein's interpretation of 1905 concerning the relationship between mass and energy {\em was flawed}. The reason for this is actually not difficult to understand: the Lorentz transformations are not far-reaching enough to make predictions about physical quantities such as energy. Special relativity is merely a simple mathematical mapping of time and coordinates. The sensible way, then, of merging special relativity with other theories is to make it compatible with other theories, rather than to assume that the Lorentz transformations have an unlimited power of prediction. This, basically, is the misconception that prevailed for the past 100 years.

\pagebreak

{\large\bf Acknowledgements:}\\
\\
The author is indebted to the following colleagues for raising the important questions discussed in this paper: Prof. Bernard Jaulin, Maison des Sciences (France); Dr. Cyrille de Souza, Maison des Sciences (France); Dr. Christopher Peters, New Jersey Inst. of Technology; Dr. Walter Becker, retired and Prof. V. Arunasalam, retired. The author sincerely thanks those individuals for the numerous hours of discussions about the topic of mass-energy equivalence.\\
\\
\\
{\large\bf Appendix:}\\
\\
The new ideas introduced in Sec.2 non-withstanding, it is quite simple to show that the total energy $H=mv^2$ transforms properly under the ``plain" Lorentz transformations. When we write the transformations in matrix form

\begin{equation}
\left( \begin{array}{c}
x^\prime\\
y^\prime\\
z^\prime\\
it^\prime
\end{array} \right)
=
\left( \begin{array}{cccc}
\gamma & 0 & 0 & i \gamma v\\
0 & 1 & 0 & 0\\
0 & 0 & 1 & 0\\
-i \gamma v/c^2 & 0 & 0 & \gamma
\end{array} \right)
\left( \begin{array}{c}
x\\
y\\
z\\
it
\end{array} \right)
\label{230}
\end{equation}

and differentiate the coordinates with respect to the proper time $\tau$, we obtain the 4-vector $(v_x^\prime=0, 0, 0, i)$ on the l.h.s. and the 4-vector $(\gamma v_x, 0, 0, i \gamma)$ on the r.h.s. The transformations can then be written in the compact form 

\begin{equation}
\left( \begin{array}{c}
v_x^\prime\\
i
\end{array} \right)
=
\left( \begin{array}{cc}
\gamma & i \gamma v\\
-i \gamma v/c^2 & \gamma
\end{array} \right)
\left( \begin{array}{c}
\gamma v_x\\
i\gamma 
\end{array} \right)
\label{231}
\end{equation}

The physical meaning of those two simple equations is very important. The first equation, as we can see, is a velocity transformation equation. The second equation was essentially a time transformation equation. When we differentiated with respect to time, it became an equation that merely describes how the quantity $\gamma$ transforms; more clearly, since $v_x = v$, then

\begin{eqnarray}
i & = & -i \frac{v^2}{c^2} \gamma^2 + i \gamma^2, \quad \mbox{or}, \nonumber\\
1 & = & \gamma^2 (1- \frac{v^2}{c^2})
\label{232}
\end{eqnarray}

As we can see, the product is indeed equal to 1. If we therefore multiply the two equations in (\ref{231}) by {\em any} constant, then the second equation will always transform that constant into itself, by virtue of the relationship in Eq.(\ref{232}). This observation is very important. The traditional approach has been to {\em assume} that there is a ``rest energy" in the particle's frame, given by the constant $m_0 c^2$, which then becomes $\gamma m_0 c^2$ in the observer's frame by virtue of the second equation; that is, the transformation is $\gamma m_0 c^2 \leftrightarrow  m_0 c^2$. But the whole notion that a quantity such as $m_0 c^2$ is a valid physical representation of energy by no means can be deduced from the simple transformations in Eq.(\ref{231}). So where did this notion emerge from? Again, it is nothing but Einstein's old interpretation.\\
\\
Let us now examine the relationship between the total energy expression $\gamma m_0 v^2$ and the plain Lorentz transformations. As we saw, the second transformation in Eq.(\ref{231}) is -after the time is taken out- essentially a transformation of {\em any} constant, regardless of how we may formulate that constant. Now, the first transformation, however, is the one that carries a much more important physical meaning. It is a velocity transformation equation. If we multiply by $m_0$, then it becomes a momentum transformation equation. As a check, multiplying by $m_0$ (and letting $v_x = v$) gives

\begin{equation}
m_0 v^\prime = (\gamma \quad i \gamma v) 
\left( \begin{array}{c}
\gamma m_0 v\\
i \gamma m_0
\end{array} \right)
\label{233}
\end{equation}

or,

\begin{equation}
m_0 v^\prime = \gamma^2 m_0 v - \gamma^2 m_0 v  = 0.
\label{234}
\end{equation}

Hence, we get the expected transformation of the relativistic momentum $\gamma m_0 v$ in the observer's frame to 0 in the particle's frame. It is also easy to verify that if we square both sides of Eq.(\ref{233}) and divide by $m_0$, we get the following equation

\begin{eqnarray}
m_0 v^{\prime 2} & = & (\gamma^3 - \gamma^3 \quad i \gamma^3 v - i \gamma^3 v) 
\left( \begin{array}{c}
\gamma m_0 v^2\\
i \gamma m_0 v
\end{array} \right)\nonumber\\
                 & = & (0 \quad 0)
\left( \begin{array}{c}
\gamma m_0 v^2\\
i \gamma m_0 v
\end{array} \right)
\label{235}
\end{eqnarray}

Hence, the transformation $\gamma m_0 v^2 \rightarrow m_0 v^{\prime 2}$ also results in a null quantity in the particle's frame. This confirms that $H= \gamma m_0 v^2$ transforms properly under the plain transformations. Note, however, that we have relied on the first transformation, not the second. As we pointed out, that second transformation -after the removal of the time-  essentially transforms any constant into itself. The association of the constant $m_0 c^2$ with the second transformation (and calling that constant ``energy") was not based on any physical concept other than Einstein's original interpretation.

\end{document}